\documentclass{article}
\usepackage{natbib}
\usepackage[utf8]{inputenc}
\usepackage{amsmath}
\usepackage{graphicx}
\usepackage{caption}
\usepackage{subcaption}
\usepackage{algpseudocode}
\usepackage{algorithm}
\usepackage{dsfont}
\usepackage{amsfonts}
\usepackage{textcomp}
\usepackage{subcaption} 
\usepackage{framed,enumitem}
\usepackage{authblk}
\usepackage[format=plain,
            labelfont={it},
            textfont=it]{caption}
\usepackage[hypertexnames=false]{hyperref}
\usepackage{cleveref} 
\newcommand{\FC}{\widetilde C}
\def\keywords{\vspace{.5em}
{\textbf{Keywords}:\,\relax%
}}

\title{Sampling by Divergence Minimization}
\author[1]{Ameer Dharamshi\thanks{Corresponding Author (e-mail: ameer.dharamshi@mail.utoronto.ca)}\thanks{ORCID iD: 0000-0002-5505-4765}}
\author[1]{Vivian Ngo}
\author[1]{Jeffrey S. Rosenthal}
\affil[1]{Department of Statistical Sciences, University of Toronto, Toronto, ON, Canada}

\date{\today}

\graphicspath{{Figures/}}
 
\begin{document}

\maketitle

\begin{abstract}

    We introduce a Markov Chain Monte Carlo (MCMC) method that is designed to sample from target distributions with irregular geometry using an adaptive scheme. In cases where targets exhibit non-Gaussian behaviour, we propose that adaption should be regional rather than global. Our algorithm minimizes the information projection component of the Kullback-Leibler (KL) divergence between the proposal and target distributions to encourage proposals that are distributed similarly to the regional geometry of the target. Unlike traditional adaptive MCMC, this procedure rapidly adapts to the geometry of the target's current position as it explores the surrounding space without the need for many preexisting samples. The divergence minimization algorithms are tested on target distributions with irregularly shaped modes and we provide results demonstrating the effectiveness of our methods.
\end{abstract}


\textbf{Declarations:}

Funding: No funding was received to assist with the preparation of this manuscript.

Conflicts of Interest/Competing Interests: The authors have no conflicts of interest to declare that are relevant to the content of this article.

Availability of data and material: Not applicable.

Code availability: The implementations of the algorithms discussed here along with all code used to generate examples can be accessed at \url{https://github.com/AmeerD/Scout-MCMC}

\tableofcontents

\newpage

\section{Introduction}

Markov Chain Monte Carlo (MCMC) is a class of algorithms designed to efficiently and effectively sample from a diverse set of target distributions \citep{mcmchandbook}. Classical MCMC methods perform excellently when the target is well-behaved and unimodal. However, when targets exhibit unusual geometry or have multiple modes, core techniques such as random walk Metropolis (RWM) tend to perform poorly. These are the challenges that motivate much active MCMC research. In this paper, we propose an algorithm that specifically aims to effectively sample from irregular and non-Gaussian target distributions. 

For targets with atypical geometry, adaptive MCMC has proven to outperform classical MCMC \citep{adaptiveMCMC, haario2001, tutorialAdaptiveMCMC}. One of the core ideas driving adaptive MCMC is that a proposal distribution that is similar in shape to the target distribution will produce higher quality samples than a generic proposal. In adaptive Random Walk Metropolis (aRWM), this is accomplished by proposing with the empirical covariance matrix of the samples produced up to the current iteration. Thus, the proposals improve as the algorithm progresses until, eventually, the empirical covariance matrix approaches the hypothetical global optimal sampler. Convergence to the target distribution can be upheld using the principles of containment and diminishing adaptation, or finite adaption \citep{ roberts_rosenthal_2007, optimalproposals}. However, aRWM does have its limitations. When the target distribution exhibits highly irregular, non-Gaussian geometry, aRWM may not perform well because a single optimal Gaussian proposal distribution as used by aRWM may not reflect the local geometry in all regions of the target distribution.

In this work, we expand on the idea that a global optimal proposal distribution may not be effective and we instead discuss the idea of region-specific sampling. We introduce an algorithm designed to sample effectively from unusual geometries by exploiting local information about the target distribution. Instead of waiting for samples to be produced in order to trigger adaptation, we use ideas from the recently popular stochastic variational inference class of methods \citep{var_inference_mcmc}. By measuring the similarity between the target and proposal distributions using the Kullback-Leibler (KL) Divergence, we use gradients to devise an update rule that is not reliant on having an existing large batch of samples. 

More specifically, consider a target distribution $p$ and a family of proposal distributions to be $q \in \mathcal{Q}$ \citep{KL_properties}. Noting that the KL Divergence between two distributions is asymmetric, we specifically focus on the I-projection of the KL Divergence, defined as  $\mathcal{D}(q||p)=E_q\left[\log\frac{q}{p}\right]$. The KL Divergence has two sides, the I- and M- projections, but we rely on the I-projection as it tends to produce an underdispersed $q$ that locks onto a specific mode of $p$, compared to the M-projection which tends to overestimate the support of $q$ \citep{MLBook}. In other words, through minimization, the I-projection produces a distribution similar to the local geometry of $p$. Such a distribution is equipped to rapidly produce samples from oddly shaped regions of a target distribution. We term this approach the Divergence Minimization (DM) Sampler.

In addition, the DM sampler is inherently modular and can be easily integrated into other MCMC methods. As an example, we implement the DM sampler as the non-tempered chain of a two-chain parallel tempering algorithm, which we call Scout MCMC. Recall that parallel tempering executes multiple chains simultaneously on the target with different levels of tempering, then randomly swaps positions of the chains to improve mode discovery \citep{replica, geyer1991markov}. The DM sampler component of Scout MCMC improves on the basic parallel tempering procedure by offering immediate adaption following swap moves.



Finally, we recognize that at each iteration, the covariance matrix produced by the gradient update rule represents a proposal distribution adept at sampling from its local region. This generated proposal distribution can reasonably be used for nearby points, assuming some degree of continuity. Thus, we introduce a two-stage extension to the DM sampler and Scout MCMC. In the first stage, we gather proposal distributions using the DM sampler or Scout MCMC. Next, we use these proposal distributions to characterize a non-adaptive Metropolis-Hastings algorithm.

Before discussing the specifics of the algorithms in Section \ref{IPJ} and Section \ref{SMCMC}, we first discuss a number of relevant related works.

\subsection{Related Works}

The DM sampler draws inspiration from \cite{NEURIPS2019_62456714}. In this paper, the authors optimize an objective function composed of the product of the entropy function and the average proposal acceptance rate. The proposals for the adaptive MCMC algorithm are then based off of gradient updates that aim to maximize this function, producing a wide range of proposals that maintain a balance between entropy and acceptance rate. In addition to supporting significant adaptation at early stages of a chain, Gradient-based Adaptive MCMC also allows for adaptation upon rejecting a proposal, a noteworthy feature as most adaptive algorithms do not directly consider the information offered by rejected samples. 

While the entropy function is a general function applied to the entire distribution, the algorithm that we present in this paper is based on the premise that, in cases with difficult geometry, it is necessary to focus on specific local regions instead of the entire target distribution when attempting to sample from the target distribution. This is accomplished by leveraging the I-Projection of the target over the set of proposal distributions. The I-projection underestimates the support of the target distribution and will hone in on one area as opposed to the entropy function which attempts to discover a range of samples from the entire target function at once \citep{journals/bstj/Shannon48, MLBook}. This regional behaviour helps to overcome limitations in the Gaussian function class typically used for proposal distributions by reducing the current region of interest into manageable pieces. 

Parallel tempering is another popular method, and is related to our Scout MCMC algorithm. In parallel tempering, multiple chains are run simultaneously on the target distribution with different levels of tempering applied. The intuition behind parallel tempering is that in the highly tempered chains, it will be easier to cross low-probability boundaries which can subsequently be randomly swapped with the non-tempered chain for mixing \citep{replica, geyer1991markov}. However, parallel tempering does have its limitations. From a computational perspective, executing many chains but ultimately only using the samples from the non-tempered chain is burdensome. 


Using parallel chains for similar purposes, in \cite{LearnFromThyNeighbour}, the authors introduce the algorithm Inter-chain Adaptation (INCA), which uses multiple stages of sampling. The first stage involves sampling the state space with parallel chains to partition the state space, while the second stage uses these predetermined regions as a guide to sample from the target distribution. The acceptance probabilities of new proposed points then depend on the region in which the current and proposed points reside. 


Finally, the Jumping Adaptive Multimodal Sampler (JAMS) algorithm addresses the challenges of multimodal sampling by front loading the computational burden of mode discovery, using optimization techniques to search for modes and subsequently incorporating this information into the sampling phase \citep{pompe2019framework}. In the sampling phase, dedicated "jump moves" are used to move between modes directly. Once in a mode, any sampler adept at unimodal sampling can be employed. Similar to how we incorporate the DM sampler into parallel tempering to form Scout MCMC, one could envision an algorithm that links the DM sampler to the JAMS algorithm to address both the irregular geometry and the multimodal sampling challenges at once.



\section{Divergence Minimization Sampler} \label{IPJ}

We propose that the challenge of sampling from irregular geometries can be overcome by focusing on smaller regions of a given target distribution that are more simple and can be adequately sampled from using common proposal distributions such as the Gaussian proposal. This region-specific sampling scheme requires addressing two core issues: identifying regions of interest and determining how best to sample from these regions. At the most granular level, each individual point in the space could constitute its own region. The rationale is that every point has its own unique surrounding geometry and thus there exists some optimal way to generate a new sample when starting at each and every point. 

The latter challenge characterizes the problem of identifying this optimal sampling procedure. To address this issue, we propose using the I-projection component of the KL Divergence as a similarity measure between the target and proposal distributions to construct proposals with similar geometry to the region around the current point \citep{MLBook}. Defining the target distribution as $p$ and the family of proposal distributions to be $q \in \mathcal{Q}$, the I-projection is $\mathcal{D}(q||p)=E_q\left[\log\frac{q}{p}\right]$. In the context of an MCMC proposal, we consider the family of proposal distributions to be Gaussian and the objective is to determine the covariance matrix that characterizes the Gaussian with minimal divergence from the target distribution at the current point. Such a proposal can be defined as:

$$
q(y|x) \sim N\left(x,LL^T\right)
$$
where $x$ is the current position, $y = x + L\epsilon$ is the proposal, $\epsilon \sim N(0,1)$ and $L$ is the Cholesky factor of the proposal covariance matrix \citep{cholesky}.

\subsection{Objective} \label{Obj}

To find a proposal distribution that minimizes the divergence with the local geometry of the target distribution, we consider using gradient updates performed at each iteration of the MCMC chain. Meanwhile, we must be cognizant of the acceptance rate. In essence, we want to have both a small I-projection so that the proposal and the target are similar, as well as a reasonably high acceptance rate so that we are able to use the samples from our proposals. As such, we propose the following as an objective function that balances both the exponential of the negative I-projection and the average acceptance rate of the proposal:

$$
s(x)=\exp\left[-\beta \mathcal{D}(q||p)\right] \cdot \int \alpha(x,y;L)q(y|x)dy
$$

In the above, $\beta$ is a hyperparameter that balances the impact of the I-projection with the average Metropolis acceptance rate defined by:
$$
\alpha(x,y;L) = \min \left\{1, \frac{p(y)}{p(x)} \right\}
$$
where $x$ is the current position, $y$ is the proposal, and $L$ is the proposal distribution Cholesky factor \citep{mcmchandbook}. Notice the negative inside the exponential term of $s(x)$. As the I-projection is non-negative, the negative exponent bounds the exponential term between 0 and 1 with the maximum obtained when $\mathcal{D}(q||p)=0$. Also note that the average acceptance rate ranges between 0 and 1. As a result of these bounds, $s \in [0,1]$ and is maximized when we have high acceptance rates with a proposal that is similar to the target. Thus, the problem of identifying a suitable proposal distribution has been reduced to maximizing $s(x)$ where the optimal proposal distribution at any given $x$ can be characterized by the corresponding optimal Cholesky factor $L_x$ at the global optimum.

To make the objective function easier to manipulate, instead of optimizing $s(x)$, we can optimize the logarithm of $s(x)$. That is: 

$$
\begin{aligned}
\log s(x) &= -\beta D(q||p) + \log \int \alpha(x,y;L)q(y|x)dy \\
&= -\beta E_q\left[\log\frac{q(y|x)}{p(y)}\right] + \log E_q\left[\alpha(x,y;L)\right] \\
&= \beta E_q\left[-\log q(y|x)\right] + \beta E_q\left[\log p(y)\right] + \log E_q\left[\alpha(x,y;L)\right] \\
&= \beta H_q + \beta E_q\left[\log p(y)\right] + \log E_q\left[\alpha(x,y;L)\right]
\end{aligned}
$$

The above statement of $\log s(x)$ contains expectations entangled with both the $p$ and $q$ distributions that precludes a closed form solution. In particular, notice that the final term is the logarithm of an expectation. Such a term is certainly not ideal for optimization purposes. The most advisable path forward to maximize $\log s(x)$ is to instead bound it below using Jensen's inequality. We can then optimize the lower bound instead of the objective directly. Thus we have:

$$
\begin{aligned}
\log s(x) &\ge \beta H_q + \beta E_q\left[\log p(y)\right] +  E_q\left[\log\alpha(x,y;L)\right] \\
&= \beta H_q + \beta E_q\left[\log p(y)\right] +  E_q\left[\log\min \left\{1, \frac{p(y)}{p(x)} \right\}\right] \\
&= \beta H_q + \beta E_q\left[\log p(y)\right] + E_q\left[\min \left\{0, \log p(y) - \log p(x) \right\}\right] \\
&= \beta H_q + \beta E_\epsilon\left[\log p(x + L\epsilon)\right] + E_\epsilon\left[\min \left\{0, \log p(x + L\epsilon) - \log p(x) \right\}\right] \\
&=: \mathcal{J}(x)
\end{aligned}
$$

$\mathcal{J}(x)$ can be used as a lower bound for the objective function and for optimization. However, while $\mathcal{J}(x)$ is certainly simpler than $\log s(x)$, a general closed form solution of the maximum at each value of $x$ is not attainable. Instead, we turn to iterative optimization methods. We choose gradient ascent as a generally accessible method to maximize $\mathcal{J}(x)$.

Gradient ascent requires specifying the gradient of $\mathcal{J}(x)$ with respect to the Cholesky factor $L$. We leave the detailed derivation of the approximate gradient of $\mathcal{J}(x)$ to Appendix \ref{Gradient} and present the final result here:

$$
\begin{aligned}
\nabla_L\mathcal{J}(x) &= \beta \text{diag}\left(\frac{1}{L_{11}},\dots,\frac{1}{L_{kk}}\right) + \frac{1}{J}\sum_{j=1}^J\frac{\beta}{p(x + L\epsilon_j)}p'(x + L\epsilon_j)\epsilon_j^T + \\
&\quad \frac{1}{J}\sum_{j=1}^J\nabla_L\min \left\{0, \log p(x + L\epsilon_j) - \log p(x) \right\}.
\end{aligned} 
$$
where $x$ is the current position, $L$ is the current value of the Cholesky factor, and $\epsilon_j$ are a sample of $J$ standard normal values used to approximate the gradients of the expectations found in $\mathcal{J}$. 

Note further that the interior of the second summation in $\nabla_L\mathcal{J}(x)$ reduces into the following two cases depending on the value of $\epsilon_j$,

$$
\begin{aligned}
&\quad\nabla_L\min \left\{0, \log p(x + L\epsilon_j) - \log p(x) \right\} \\
&=\begin{cases} 0&\mbox{if } \log p(x + L\epsilon_j) \ge \log p(x) \\
\frac{1}{p(x + L\epsilon_j)}p'(x + L\epsilon_j)\epsilon_j^T & \mbox{if } \log p(x + L\epsilon_j) < \log p(x) \end{cases}.
\end{aligned}
$$

The above gradient characterizes the gradient update rule $L_{t+1} = L_t + \gamma\nabla_L\mathcal{J}(x)$ where $t$ is the time step of gradient ascent and $\gamma$ is the step size used to maximize $\mathcal{J}(x)$. Here we make the practical note that due to the presence of the $p(x+L\epsilon_j)^{-1}$ term in the gradient, $\epsilon_j$ values that result in proposals with negligible density can cause an explosion of the gradient. We thus set a large threshold value of $h$ to catch elements in the gradient matrix with absolute values greater than $h$, and set the offending values to $\pm h$ respectively. This event is rare in practice but more common in tail geometries where the fraction of potentially offending proposals is higher. 

Now, if we use this procedure to identify a value of $L$ to maximize $\mathcal{J}(x)$ for each point $x$, call these $L_x$, we could then characterize a Metropolis-Hastings algorithm using these Cholesky factors.

However, we recognize that a great number of steps would be necessary to optimize $\mathcal{J}(x)$ to within some small error threshold. As gradient updates can be computationally expensive, executing a complete run of gradient ascent at every iteration of an MCMC algorithm would be untenable.


We propose that instead of fully optimizing $\mathcal{J}(x)$ at every iteration, a process that requires many expensive steps, we perform one step of gradient ascent at every MCMC iteration. This will provide approximations of the point-wise optimal sampler discussed so far with the following justifications. First, we note that the early steps of gradient ascent tend to be the most influential and thus a complete run of gradient ascent is not absolutely necessary. Secondly, in practical contexts, changes in geometry are typically gradual which implies that nearby points experience similar behaviour, and by extension, similar gradients. While the proposal distribution is not fully optimized at every iteration, on aggregate, the proposal distributions become more optimal as iterations progress.

\subsection{Algorithm Details}

We now gather the results of the above discussions into a complete algorithm summary. The Divergence Minimization sampler's objective function and gradient update rule produce a series of covariance matrices for generating Gaussian proposals with an MCMC framework. Consistent with the acceptance rule in the objective function, we incorporate a Metropolis rule for proposal acceptance. At each iteration, we accept the proposal $y$ from the current position $x_t$ with probability:
$$
\alpha(x_t,y|C_t) = \min\left\{1, 
\frac{p(y)}{p(x_t)}\right\}
$$
where $t$ is the current MCMC iteration, and $C_t$ is the current Cholesky factor of the proposal distribution's covariance matrix. Note that $C_t$ represents the partially optimized Cholesky factor as opposed to the fully optimized $L_x$ used previously. We reserve discussion of convergence issues for Section \ref{Convergence}.

\begin{algorithm} 
\caption{Divergence Minimization Sampler with Perpetual Adaption}
\label{IPalg}
\begin{algorithmic}[1]
\State \textbf{Inputs (defaults)}: target $p(x)$, balancing parameter $\beta$ (0.2), initial point $x_0$, step size $\gamma$ (0.002), step threshold $h$ (10/$\gamma$), initial scaling $\sigma$ (2), iterations $M$
\State \textbf{Initialize}: $C_0$ := $\sigma\mathds{1}$
\For {t = 0,...,M}
\State Generate $\epsilon_t \sim N(\mathbf{0}, \mathds{1})$
\State Propose $y = x_t + C_t\epsilon_t$
\State Compute $G = \nabla_L\mathcal{J}(x_t)$
\State Accept $y$ with probability $\alpha(x_t,y|C_t)$
\State Update $x_{t+1} = y$ if accepted or $x_{t+1} = x_t$ if rejected.
\State If element $|G_{ij}| > h$, set $G_{ij} = \text{sign}(G)\cdot h$
\State Update Cholesky Factor: $C_{t+1} \leftarrow C_t + \gamma G$
\EndFor
\end{algorithmic}
\end{algorithm}


Algorithm \ref{IPalg} summarises the DM sampler with perpetual adaption. In its most basic form, the initial Cholesky factor is set to a diagonal matrix with equal scaling along each dimension though a more complex initialisation would also be valid. Furthermore, parameters including the step size and balancing parameters are constant and supplied as inputs although they could be adapted along with the Cholesky factor. 

\subsection{Divergence Minimization: A Case Study} \label{dmexample}

\begin{figure}[ht]
    \centering
    \includegraphics[height=4cm]{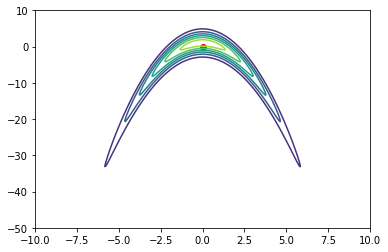}
    \caption{Banana distribution contours. (Note: lighter contours indicate higher density, red dot indicates the origin)}
    \label{fig:banana}
\end{figure}

To understand the behaviour of the DM sampler, we examine a case study using a single banana distribution. The banana distribution is a unimodal distribution with non-Gaussian contours. For context, the contours of this distribution are presented in Figure \ref{fig:banana}. The banana distribution is known to be a difficult distribution to sample from with basic MCMC algorithms because of its quickly changing local geometry, especially in the two tails \citep{banana1, haario2001}. 

\begin{figure}[ht!]

\begin{subfigure}{0.49\textwidth}
\includegraphics[width=0.9\linewidth, height=4cm]{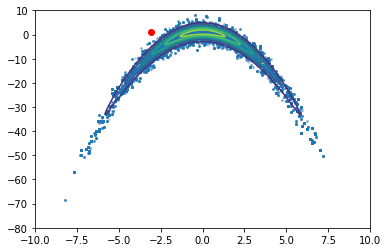} 
\caption{aRWM}
\label{fig:bananasamps1}
\end{subfigure}
\begin{subfigure}{0.49\textwidth}
\includegraphics[width=0.9\linewidth, height=4cm]{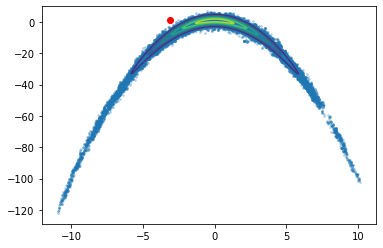}
\caption{DM sampler}
\label{fig:bananasamps2}
\end{subfigure}

\caption{Banana distribution samples. The aRWM samples are more sparse and there are gaps in the tails whereas the DM sampler produces more samples in the tails that reach further outwards. (Note: Red dot indicates starting points and blue dots indicate samples)}
\label{fig:bananasamps}
\end{figure}

An intuitive way to understand the behaviour of the DM sampler is to examine its samples. Figure \ref{fig:bananasamps} presents parallel results from an adaptive Random Walk Metropolis (aRWM) and DM sampler run. The aRWM algorithm used in this case study and subsequent examples is described by \cite{ExamplesAdaptiveMCMC}. Each algorithm was run for 20,000 iterations with the first 1,000 removed as burn-in. Visually, we notice in the DM sampler results in Figure \ref{fig:bananasamps2} that the interior of the contours is evenly explored whereas aRWM has blank gaps within the tails of the contours. 

Quantitatively, we compare the algorithms using the acceptance rate and the expected squared jumping distance (ESJD). The ESJD used here balances the goals of a high acceptance rate with the increased exploration of larger steps and is defined as $ESJD = \sum_{t=2}^M ||x_t - x_{t-1}||_2^2$ \citep{esjd_paper, optimal_scaling_esjd}. The DM sampler produced an acceptance rate of 71.43\% with an ESJD of 2.1 as compared to the 8.45\% acceptance rate and ESJD of 8.4 of aRWM. Since we know that the DM sampler has a higher acceptance rate, this suggests that the DM sampler takes smaller steps and is perhaps less efficient in terms of exploration than aRWM. With that said, more careful steps suggest a lower risk of missing regions of interest.

\begin{figure}[ht]
    \centering
    \includegraphics[height=4cm]{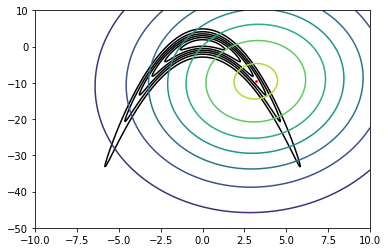}
    \caption{Proposal distribution contours from the final iteration of aRWM centred at the final sample imposed on the banana distribution contours. Notice that the contours do not match the behaviour of the target distribution.}
    \label{fig:arwmcontour}
\end{figure}

Such behaviours can be explained by examining the contours of the proposal distribution at different points of the target distribution. Figure \ref{fig:arwmcontour} presents the contours of the final proposal distribution of the aRWM run centered at the final sample. In other words, they are the contours of the covariance matrix of all samples generated. Notice that the contours have largely failed to adapt to the specific geometry of the target distribution. They have simply expanded so that all regions of meaningful density in the target are covered by the proposal distribution at any given time but have not conformed to the unique geometry of the target distribution \citep{B_dard_2007}. One might expect that we could achieve similar success with even a simple RWM algorithm, given a large enough proposal distribution.

\begin{figure}[h]

\begin{subfigure}{0.49\textwidth}
\includegraphics[width=0.9\linewidth, height=4cm]{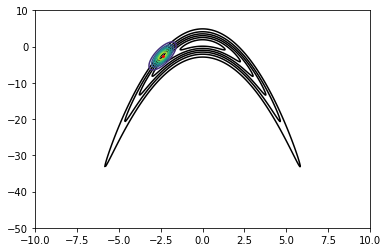} 
\caption{Left tail contour}
\label{fig:left}
\end{subfigure}
\begin{subfigure}{0.49\textwidth}
\includegraphics[width=0.9\linewidth, height=4cm]{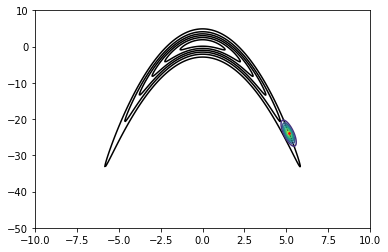}
\caption{Right tail contour}
\label{fig:right}
\end{subfigure}

\caption{Sample DM sampler contours from the same algorithm execution. Notice that the proposal contours in the given iterations conform to the local geometry of the target distribution.}
\label{fig:iprojcontour}
\end{figure}

We contrast this behaviour to that demonstrated by the DM sampler in Figure \ref{fig:iprojcontour}. The DM sampler delivers on the promise of adaptation to local behaviour as illustrated by the contours closely matching the region of interest. The proposal distributions benefit from adapted covariance matrices that align with the current tail, resulting in a dramatically reduced likelihood of bad proposals as compared to the aRWM proposals.


In summary, the contour plots demonstrate the intended behaviour of the DM sampler to adapt to local regions of interest. Such behaviour aligns with the objective of producing desirable adaptation. While it is common for algorithms to simply adapt to the \textit{scale} of a target distribution, the DM sampler adapts to the \textit{behaviour} of the target distribution, a completely different and much more challenging task that is especially useful for target distributions with unique geometry. Furthermore, while in this instance it seems that aRWM outperforms in efficiency, we must question whether adapting to just the \textit{scale} of a target distribution is scalable in higher dimensions given that the density will become more and more sparse. 

\subsection{Convergence and Finite Adaptation} \label{Convergence}

In a standard adaptive scheme, the algorithm typically involves certain technical conditions (such as diminishing adaptation and containment, or finite adaption) to guarantee convergence to the target distribution \citep{roberts_rosenthal_2007, optimalproposals}. In this work, we argue that certain target distributions, such as those with unusual geometry or those with many unique modes, lend themselves to perpetual adaptation as no single Gaussian proposal distribution could hope to sample well in all regions of interest. The banana example in the previous section is a good example to illustrate this. As the banana distribution is clearly non-Gaussian, a single non-adapting Gaussian proposal cannot appropriately orient itself in the apex \textit{and} in both tails. However, the consequence of embracing perpetual adaptation is that the standard convergence framework for adaptive MCMC is no longer compatible. 

Our goal is to sample efficiently using region-specific proposal distributions while still fulfilling the requirements of the standard convergence framework. As such, we propose a two phase approach that limits adaptation to a finite number of iterations and subsequently transfers the lessons learned in adaptation to a Metropolis-Hastings framework. By limiting adaption to a finite number of iterations, convergence of the non-adaptive phase to the target distribution is guaranteed \citep{roberts_rosenthal_2007, optimalproposals}.

Recall that the basis of the DM sampler is to approximate the optimal proposal distribution characterized by the Cholesky factor $L_x$ that maximizes the objective function $s(x)$. As discussed in Section \ref{Obj}, if we knew the values of all $L_x$, we could produce a simple Metropolis-Hastings algorithm with defined proposal distributions. Of course, we have seen that optimizing $s(x)$ is difficult for a single point, let alone all points in space. Fortunately, the procedure described in Algorithm \ref{IPalg} constructs Cholesky factors $C_t$ at each iteration $t$ to approximate the given location's optimal proposal structure for sampling. If we record these Cholesky factors after each iteration, they can act as a proxy for the optimal Cholesky factor for nearby points as well, assuming some degree of continuity. Thus, after generating a collection of points and their associated Cholesky factors in the adaptive phase, in each iteration of the non-adaptive phase we select the Cholesky factor associated with the closest adaptive phase sample to construct a proposal covariance matrix for the current iteration. The algorithm thus proposes points from the distribution $q(y|x_t) \sim N(x_t,C_t C_t^T)$ and accept with the following rule:
$$
\alpha_f(x_t,y|C_t,\FC_y) = \min\left\{1,\frac{p(y)q(x_t|y)}{p(x_t)q(y|x_t)}\right\}
$$
where $x_t$ is the current position, $y$ is the proposal, $q(y|x_t) \sim N(x_t,C_t C_t^T)$, $q(x_t|y) \sim N(y,\FC_y \FC_y^T)$, and $C_t$ and $\FC_y$ are the Cholesky factors from the adaptive phase iterations that correspond to the points closest to $x_t$ and $y$ respectively. In other words, instead of calculating a new Cholesky factor for every new point, we select the point from our adaptive phase that is closest to the new point and use its corresponding (approximate) Cholesky factor. This non-adaptive phase adheres to the standard validity criteria of a non-adaptive Metropolis-Hastings algorithm. A complete algorithm summary of this scheme is presented in Algorithm \ref{finadapt}. 

\begin{figure}[ht]
    \centering
    \includegraphics[height=4cm]{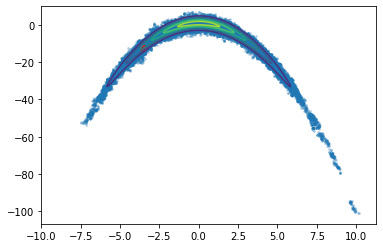}
    \caption{Finite adaptation DM sampler variant. Samples presented only include those from the non-adaptive phase.}
    \label{fig:finite}
\end{figure}

We test the finite adaptation variant of the DM sampler on the banana distribution presented in Section \ref{dmexample}. Essentially, once the adaptive phase completes, we consolidate the samples and covariance matrices and then begin the non-adaptive phase at the last adaptive phase sample. In Figure \ref{fig:finite}, we present 20,000 samples generated from the non-adaptive phase. In addition to the visual indication of the non-adaptive samples covering the relevant portions of the state space, we note that the acceptance rate is 60.62\%, the proportion of samples in the left side of the distribution is 51.7\%, and the sample mean of $\begin{bmatrix} -0.27 & -6.73\end{bmatrix}$ is approaching the true mean. These diagnostics indicate the algorithm is sampling well and is converging to the target distribution as expected. 

We note that the finite adaption version of the DM sampler performs similarly to the perpetually adapting version with the added benefit of adhering to established convergence criteria.

\begin{algorithm}[h]
\caption{Divergence Minimization Sampler with Finite Adaptation}
\label{finadapt}
\begin{algorithmic}[1]
\State \textbf{Inputs (defaults)}: target $p(x)$, balancing parameter $\beta$ (0.2), initial point $x_0$, step size $\gamma$ (0.002), step threshold $h$ (10/$\gamma$), initial scaling $\sigma$ (2), iterations $M$, finite adaptation threshold $F$ ($M$/2), finite subsample size $s$ ($M$/20)
\State \textbf{Initialize}: $C_0$ := $\sigma\mathds{1}$
\State \textbf{Adaptive Phase}
\For {t = 0,...,F}
\State Generate $\epsilon_t \sim N(\mathbf{0}, \mathds{1})$
\State Propose $y = x_t + C_t\epsilon_t$
\State Compute $G = \nabla_L\mathcal{J}(x_t)$
\State Accept $y$ with probability $\alpha(x_t,y|C_t)$
\State Update $x_{t+1} = y$ if accepted or $x_{t+1} = x_t$ if rejected.
\State If element $|G_{ij}| > h$, set $G_{ij} = \text{sign}(G)\cdot h$
\State Update Cholesky Factor: $C_{t+1} \leftarrow C_t + \gamma G$
\EndFor
\State Let $S$ be a sample of $s$ points from ${0,1,...F}$
\State \textbf{Non-Adaptive Phase}
\For {t = F+1,...,M}
\State Select $C_t := C_{i}$ where $i\in S$, such that $d(x_i,x_t)$ is minimized.
\State Generate $\epsilon_t \sim N(\mathbf{0}, \mathds{1})$
\State Propose $y_t = x_t + C_t\epsilon_t$
\State Select $\FC_y := C_{j}$ where $j\in S$, such that $d(x_j,y)$ is minimized.
\State Accept $y$ with probability $\alpha_f(x_t,y|C_t,\FC_y)$
\State Update $x_{t+1} = y$ if accepted or $x_{t+1} = x_t$ if rejected.
\EndFor
\end{algorithmic}
\end{algorithm}




\bigskip

\noindent\textbf{Remark}: We  now comment on the choice of the Metropolis acceptance rule for the original DM sampler as well as discuss an alternative that could perhaps motivate future work. Recall that each iteration of the adaptive phase triggers the gradient update rule. Any proposal under this framework will be asymmetric which at first glance would suggest the use of a Metropolis-Hastings acceptance rule \citep{Hastings}. Suppose for a moment that we were to consider a Metropolis-Hastings rule. In other words, we replace the acceptance rule $\alpha$ with the following: 

$$
\alpha^*(x_t,y|C_t) = \min\left\{1,\frac{p(y)q(x_t|y)}{p(x_t)q(y|x_t)}\right\}
$$
where $x_t$ is the current position, $y$ is the proposal, $C_t$ is the Cholesky factor of the proposal covariance matrix, $q(y|x_t) \sim N(x_t,C_t C_t^T)$, and $q(x_t|y) \sim N(y,(C_t+\gamma\nabla_L\mathcal{J}(x_t))(C_t+\gamma\nabla_L\mathcal{J}(x_t))^T)$. The distribution of $q(x_t|y)$ considers the gradient step made in the process of moving from $x_t$ to $y$, reflecting the asymmetry involved in returning from $y$ to $x_t$. In this case, reversibility would be upheld at each individual iteration without introducing any finite adaptation \citep{ROBERTS1994207, containmentAdaptiveMCMC, Craiu_2015}. However, the proposal kernels across iterations are not necessarily identical. Concretely, visiting, leaving, and then returning to a point can result in different proposal kernels at the same point due to the use of only a single gradient step at each iteration. Thus, each individual step under the hypothetical Metropolis-Hastings setup would be reversible but, in aggregate, the entire chain may not be. This perpetual adaption represents a departure from the established convergence theory. In this paper, we have instead decided to proceed with a finite adaptation scheme that does guarantee convergence to the target distribution.



\section{Scout MCMC} \label{SMCMC}


The DM sampler is designed as a general procedure for rapid adaption to local geometry. Given the self-contained setup, it can be combined with other MCMC frameworks such as parallel tempering to produce more complex samplers. In this section, we introduce an extension that combines the DM Sampler with a two-chain parallel tempering setup. Specifically, the untempered first chain uses the DM sampler, and the second chain is tempered by either a factor provided by the user or one proportional to the number of dimensions \citep{tawn2019weightpreserving}. Such an approach benefits from both the regional adaption of the DM sampler and the global exploration of parallel tempering swap moves. Given the single tempered chain searching for new regions of density, we term this approach Scout MCMC.

\subsection{Algorithm Details}

Scout MCMC generates proposals for the main chain $q(y_t|x_t) \sim N(x_t,C_t C_t^T)$, and accepts with probability: 

$$
\begin{aligned}
\alpha(x_t,y_t|C_t) &= \min\left\{1,\frac{p(y_t)}{p(x_t)}\right\}
\end{aligned}
$$
Then, it adapts the main chain Cholesky factor by $\gamma\nabla_L\mathcal{J}(x_t)$ as before. Next, a proposal for the scout chain is generated as $q(c_t|s_t) \sim N(s_t, \sigma_s\mathds{1})$, which is accepted according to the following rule: 

$$
\begin{aligned}
\alpha_s(s_t,c_t) &= \min\left\{1, \frac{p(c_t)^\tau}{p(s_t)^\tau}\right\}
\end{aligned}
$$
Finally, Scout MCMC considers swapping $x_{t+1}$ and $s_{t+1}$ every $k$ iterations
according to the swap rule:

$$
\begin{aligned}
\alpha_{\text{swap}}(x,s) &= \min\left\{1,\frac{p(x)^\tau p(s)}{p(x)p(s)^\tau}\right\} 
\end{aligned}
$$

\begin{algorithm}[h]
\caption{Scout MCMC with Perpetual Adaption}
\label{Scout}
\begin{algorithmic}[1]
\State \textbf{Inputs (defaults)}: target $p(x)$, balancing parameter $\beta$ (0.2), temperature $\tau$ (0.1), initial point $x_0$, step size $\gamma$ (0.002), step threshold $h$ (10/$\gamma$), initial scaling $\sigma$ (2), tempered scaling $\sigma_s$ (9), iterations $M$, swap frequency $k$ (20)
\State \textbf{Initialize}: $C_0$ := $\sigma\mathds{1}$, $s_0 = x_0$
\For {t = 0,...,M}
\State \textbf{Main Chain Step}
\State Generate $\epsilon_t \sim N(\mathbf{0}, \mathds{1})$
\State Propose $y_t = x_t + C_t\epsilon_t$ 
\State Compute $G = \nabla_L\mathcal{J}(x)$
\State Accept $y_t$ with probability $\alpha(x_t,y_t|C_t)$
\State Update $x_{t+1} = y_t$ if accepted or $x_{t+1} = x_t$ if rejected.
\State If element $|G_{ij}| > h$, set $G_{ij} = \text{sign}(G)\cdot h$
\State Update Cholesky Factor: $C_{t+1} \leftarrow C_t + \gamma G$
\State \textbf{Scout Step}
\State Propose $c_t \sim N(s_t, \sigma_s\mathds{1})$ and accept with probability $\min\left\{1, \frac{p(c_t)^\tau}{p(s_t)^\tau}\right\}$
\State Update $s_{t+1} = c_t$ if accepted or $s_{t+1} = s_t$ if rejected.
\State \textbf{Swap Step}
\If {$t \equiv 0 \mod{k}$}
\State Swap $x_{t+1}$ and $s_{t+1}$ with probability $\min\left\{1,\frac{p(s_{t+1})p(x_{t+1})^\tau}{p(x_{t+1})p(s_{t+1})^\tau}\right\}$
\EndIf
\EndFor
\end{algorithmic}
\end{algorithm}

Algorithm \ref{Scout} provides pseudocode for the implementation of Scout MCMC. Once again, control over step size and initial scaling is determined by the user to allow flexibility between targets. For example, depending on the expected global region of interest, the tempered chain scaling can be adjusted. Additional details can be added such as adapting the scaling of the tempered chain or varying the limit on the frequency of swap moves.

\subsection{Finite Adaptation}

Similar to the DM sampler, we present a two-phase finitely adapting variant of Scout MCMC. The first phase is the procedure presented in Algorithm \ref{Scout}. In the second, non-adaptive phase, the structure of the scout chain does not change. However, the main chain follows the same process as the finitely adapting DM sampler where the Cholesky factor corresponding to the nearest iteration of the adapting phase is used to construct proposal distributions in the non-adaptive phase. This reduces the non-adapting phase to a Metropolis-Hastings algorithm. We present the pseudocode associated with the finitely adapting Scout MCMC in Algorithm \ref{Scoutfinite}.

\begin{algorithm}
\caption{Scout MCMC with Finite Adaptation}
\label{Scoutfinite}
\begin{algorithmic}[1]
\State \textbf{Inputs (defaults)}: target $p(x)$, balancing parameter $\beta$ (0.2), temperature $\tau$ (0.1), initial point $x_0$, step size $\gamma$ (0.002), step threshold $h$ (10/$\gamma$), initial scaling $\sigma$ (2), tempered scaling $\sigma_s$ (9), iterations $M$, finite adaptation threshold $F$ ($M$/2), swap frequency $k$ (20), finite subsample size s ($M$/20)
\State \textbf{Initialize}: $C_0$ := $\sigma\mathds{1}$, $s_0 = x_0$
\State \textbf{Adaptive Phase}
\For {t = 0,...,F}
\State \textbf{Main Chain Step}
\State Generate $\epsilon_t \sim N(\mathbf{0}, \mathds{1})$
\State Propose $y_t = x_t + C_t\epsilon_t$ 
\State Compute $G = \nabla_L\mathcal{J}(x)$
\State Accept $y_t$ with probability $\alpha(x_t,y_t|C_t)$
\State Update $x_{t+1} = y_t$ if accepted or $x_{t+1} = x_t$ if rejected.
\State If element $|G_{ij}| > h$, set $G_{ij} = \text{sign}(G)\cdot h$
\State Update Cholesky Factor: $C_{t+1} \leftarrow C_t + \gamma G$
\State \textbf{Scout Step}
\State Propose $c_t \sim N(s_t, \sigma_s\mathds{1})$ 
\State Accept $c_t$ with probability $\min\left\{1, \frac{p(c_t)^\tau}{p(s_t)^\tau}\right\}$
\State Update $s_{t+1} = c_t$ if accepted or $s_{t+1} = s_t$ if rejected.
\State \textbf{Swap Step}
\If {$t \equiv 0 \mod{k}$}
\State Swap $x_{t+1}$ and $s_{t+1}$ with probability $\min\left\{1,\frac{p(s_{t+1})p(x_{t+1})^\tau}{p(x_{t+1})p(s_{t+1})^\tau}\right\}$
\EndIf
\EndFor
\State Let $S$ be a sample of $s$ points from ${0,1,...F}$
\State \textbf{Non-Adaptive Phase}
\For {t = F+1,...,M}
\State \textbf{Main Chain Step}
\State Select $C_t := C_{i}$ where $i\in S$, such that $d(x_i,x_t)$ is minimized.
\State Generate $\epsilon_t \sim N(\mathbf{0}, \mathds{1})$
\State Propose $y_t = x_t + C_t\epsilon_t$
\State Select $\FC_t := C_{j}$ where $j\in S$, such that $d(x_j,y)$ is minimized.
\State Accept $y$ with probability $\alpha_f(x_t,y|C_t,\FC_t)$
\State Update $x_{t+1} = y$ if accepted or $x_{t+1} = x_t$ if rejected.
\State \textbf{Scout Step}
\State Propose $c_t \sim N(s_t, \sigma_s\mathds{1})$ and accept with probability $\min\left\{1, \frac{p(c_t)^\tau}{p(s_t)^\tau}\right\}$
\State Update $s_{t+1} = c_t$ if accepted or $s_{t+1} = s_t$ if rejected.
\State \textbf{Swap Step}
\If {$t \equiv 0 \mod{k}$}
\State Swap $x_{t+1}$ and $s_{t+1}$ with probability $\min\left\{1,\frac{p(s_{t+1})p(x_{t+1})^\tau}{p(x_{t+1})p(s_{t+1})^\tau}\right\}$
\EndIf
\EndFor
\end{algorithmic}
\end{algorithm}

In the next section, we present examples demonstrating that similar to the DM sampler, the finite adaption version of Scout MCMC performs similarly in practice to the perpetually adapting version but has the theoretical advantage of adhering to established convergence criteria.

\section{Examples} \label{Examples}

In this section, we examine the performance of the DM Sampler and Scout MCMC using a variety of target distributions. We focus on distributions with atypical geometry as well as a subset of multimodal distributions that have contiguous modes. It is important to note that traditional diagnostics such as effective sample size (ESS) will be misleading in the case of multimodal distributions \citep{turner2017does, elvira2018rethinking}. ESS specifically may prefer a sample that fails to leave the initial mode as compared to a sample that explores modes separated by a low probability chasm. ESJD is arguably a better diagnostic as it increases with increased step size and acceptance rate, both being favourable behaviours. Recall that the ESJD is defined as $ESJD = \sum_{t=2}^M ||x_t - x_{t-1}||_2^2$.  


Given that there is a lack of consensus on appropriate diagnostics for targets with more than one mode, we have selected target distributions with easily computed true expected values to use as reference points for the simulations. Going forward, we will refer to the true expected value as $E[X]$, the estimated expected value with an MCMC sample as $\hat{E}[X]$, and the Euclidean distance between the true and estimated values as $d(E[X],\hat{E}[X])$.

As an example of a target distribution with easily computed expectations, the basis vector target that will be discussed in detail consists of a mixture of Gaussian distributions where each component Gaussian lies on one of the basis vectors and all are equidistant from the origin. This leads to a target with negligible density at the origin but with an expected value that is simply at the origin itself. A similar but much more challenging target consisting of a mixture of banana distributions presents a target with a mean at the origin that also has complex geometry. In these instances, we can use the distance from the sample mean to the origin, the true mean, to evaluate algorithm performance.

In the following examples, we compare the DM Sampler and Scout MCMC with standard Random Walk Metropolis (RWM), adaptive RWM (aRWM), Metropolis-adjusted Langevin algorithm (MALA), and Parallel Tempering (PT). For clarity, RWM generates proposals with a single shared Gaussian distribution, aRWM generates proposals using the empirical covariance matrix up to the current iteration, MALA uses gradient information to improve proposals, and PT executes multiple RWM chains on the target distribution with different levels of tempering applied \citep{replica, geyer1991markov, optimal_scaling_mala}. For consistency, we match the maximum tempering level used by parallel tempering to the level used by the scout chain in Scout MCMC. We also execute two versions of parallel tempering: one with 2 chains to match Scout MCMC, and one with 5 or 10 chains as would be more likely in practice. Finally, we include both the fully adaptive versions of the DM Sampler and Scout MCMC along with the variants that limit adaptation and transition to a second non-adaptive phase. The code used to generate the following examples along with implementations of each algorithm in Python are provided to supplement the discussion\footnote{https://github.com/AmeerD/Scout-MCMC}. 

\subsection{Double Banana Distribution}

\begin{figure}[ht]
    \centering
    \includegraphics[height=4cm]{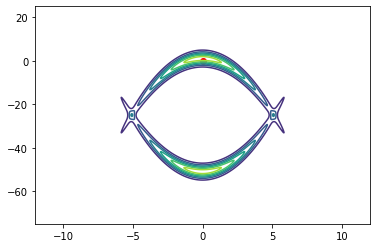}
    \caption{Double banana distribution contours. (Note:  lighter contours indicate higher density, red dot indicates the origin)}
    \label{fig:double}
\end{figure}

The first distribution we consider is an extension of the banana distribution examined in Section \ref{dmexample}. Specifically, we consider a pair of banana distributions with overlap in the tails. This results in two primary modes along the curves of the two bananas as well as two secondary modes at the intersections. Figure \ref{fig:double} provides the contours of this distribution. 

All of the algorithms are run for 50,000 iterations with the first 1,000 samples discarded as burn-in. Table \ref{tab:bans} presents the results of this experiment. For this specific distribution, the target mean is $\begin{bmatrix} 0 & -25 \end{bmatrix}$. 

Notice that the samples of aRWM and Scout MCMC are closest to the mean of the distribution. It is worth noting that there is negligible density at the mean as illustrated by Figure \ref{fig:double} so the ability to achieve the correct mean indicates that both bananas have been visited. In comparison, standard RWM, MALA, and parallel tempering have not achieved the level of success of the other algorithms. Even with 5 chains, parallel tempering has largely failed to converge within the 50,000 iterations. The distinction between aRWM and Scout MCMC lies in the efficiency diagnostics. We see that Scout MCMC accepts over 10 times as many proposals as aRWM though it has a smaller ESJD. A large acceptance rate is not necessarily indicative of a better algorithm but with both algorithms performing similarly, this could indicate that Scout MCMC produces higher quality proposals. Since ESJD is a measure of both the acceptance rate and the step size made with each move but Scout MCMC has a much higher acceptance rate, this would indicate that aRWM proposes moves with much greater step sizes than Scout MCMC. This is expected, however, as Scout MCMC uses a user-specified cooldown period where the main chain makes local moves and does not swap with the scout chain. Finally, we note that the finite adaptation variants of the DM sampler and Scout MCMC both perform similarly to their fully adapting counterparts though they tend to accept fewer proposals. 

\begin{center}
\begin{table}[ht]
\centering
 \begin{tabular}{ l | c c c c c} 
 \hline
 & Accept (\%) & $\hat{E}$[X] &  $d$(E[X],$\hat{E}$[X]) & ESJD \\ [0.5ex] 
 \hline
 RWM & 51.89 & $\begin{bmatrix} +0.31 & -17.77 \end{bmatrix}$ & 7.24 & 0.78 \\ 
 \hline
 aRWM & 7.76 & $\begin{bmatrix} +0.16 & -24.84 \end{bmatrix}$ & 0.23 & 40.2 \\
 \hline
 PT (2 chains) & 37.59 & $\begin{bmatrix} -2.29 & -19.31 \end{bmatrix}$ & 6.13 & 0.99 \\
 \hline
 PT (5 chains) & 40.25 & $\begin{bmatrix} -2.40 & -4.210 \end{bmatrix}$ & 20.92 & 2.16 \\
 \hline
 MALA & 88.54 & $\begin{bmatrix} +0.17 & -1.94 \end{bmatrix}$ & 23.06 & 0.17 \\
 \hline
 DM Sampler & 83.50 & $\begin{bmatrix} -2.35 & -23.76 \end{bmatrix}$ & 2.66 & 0.80 \\
 \hline
 DM Finite & 68.59 & $\begin{bmatrix} -2.25 & -24.02 \end{bmatrix}$ & 2.45 & 0.61 \\
 \hline
 Scout MCMC & 83.41 & $\begin{bmatrix} -0.22 & -23.78 \end{bmatrix}$ & 1.24 & 11.1 \\
 \hline
 Scout Finite & 73.57 & $\begin{bmatrix} -0.22 & -18.83 \end{bmatrix}$ & 6.17 & 11.8 \\
 [1ex] 
 \hline
\end{tabular}
\caption{Double banana target results. We see here that aRWM and Scout MCMC produced sample means that are closest to the true mean. While aRWM has a greater ESJD value, Scout MCMC has a greater acceptance rate.}
\label{tab:bans}
\vspace{-20pt}
\end{table}
\end{center}

\begin{figure}[ht]
\centering
\begin{subfigure}[t]{0.24\textwidth}
\includegraphics[width=0.9\linewidth, height=4cm]{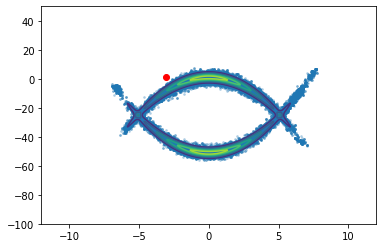} 
\caption{RWM}
\label{fig:rdb}
\end{subfigure}\hfil
\begin{subfigure}[t]{0.24\textwidth}
\includegraphics[width=0.9\linewidth, height=4cm]{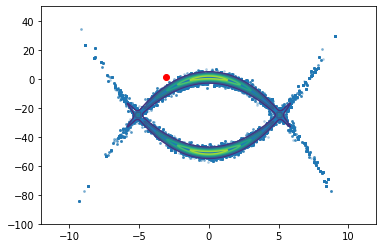}
\caption{aRWM}
\label{fig:adb}
\end{subfigure}\hfil
\begin{subfigure}[t]{0.24\textwidth}
\includegraphics[width=0.9\linewidth, height=4cm]{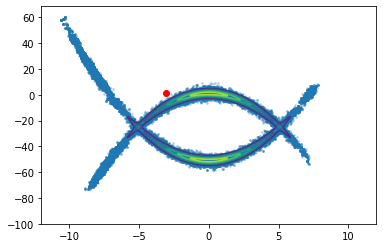}
\caption{PT (2 chains)}
\label{fig:pt2db}\hfil
\end{subfigure}
\begin{subfigure}[t]{0.24\textwidth}
\includegraphics[width=0.9\linewidth, height=4cm]{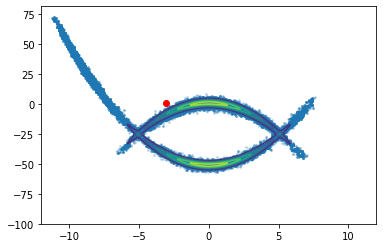}
\caption{PT (5 chains)}
\label{fig:pt5db}
\end{subfigure}

\begin{subfigure}[t]{0.32\textwidth}
\includegraphics[width=0.9\linewidth, height=4cm]{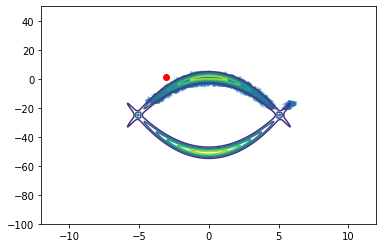} 
\caption{MALA}
\label{fig:maladb}
\end{subfigure}\hfil
\begin{subfigure}[t]{0.32\textwidth}
\includegraphics[width=0.9\linewidth, height=4cm]{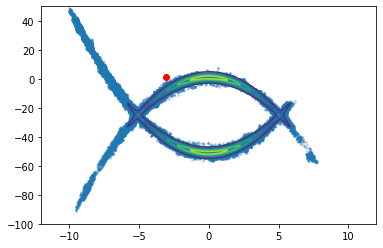} 
\caption{DM sampler}
\label{fig:idb}
\end{subfigure}\hfil
\begin{subfigure}[t]{0.32\textwidth}
\includegraphics[width=0.9\linewidth, height=4cm]{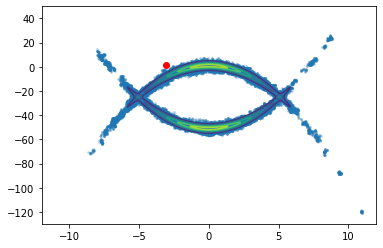}
\caption{Scout MCMC}
\label{fig:sdb}
\end{subfigure}

\caption{Double Banana Samples. PT and the DM sampler have trouble moving away from tail regions. The best performing algorithms are aRWM and Scout MCMC as they achieve the most accurate sample mean and highest ESJD values. (Note: Red dot indicates starting points and blue dots indicate samples)}
\label{fig:dsamps}
\end{figure}

In addition to sample diagnostics, we also examine the samples themselves in Figure \ref{fig:dsamps}. A notable observation is the dramatic imbalance of the parallel tempering samples in Figures \ref{fig:pt2db} and \ref{fig:pt5db} as well as the DM samples in Figure \ref{fig:idb}. RWM also experiences slight imbalance but more notably does not reach far into the tails within the number of iterations. Most surprising is that MALA has not managed to reach the bottom banana regardless of the choice of parameters. We attribute this largely to the gradient pulling proposals away from the tails and thus hampering exploration. The contrast between the DM sampler and Scout MCMC samples highlights the regulating abilities of the Scout chain to help the DM sampler escape from extreme regions. In this example, the left tail in Figure \ref{fig:idb} could be considered an extreme region. Both the aRWM and Scout MCMC plots exhibit desirable sampling behaviour as the samples are well dispersed over the target and seemingly balanced. However, the primary difference between the two algorithms in this example is the relative concentration of samples due to aRWM's tendency to reject proposals and stay at the same points whereas Scout MCMC produces a larger number of unique points. 

\begin{figure}[ht]

\begin{subfigure}{0.49\textwidth}
\includegraphics[width=0.9\linewidth, height=4cm]{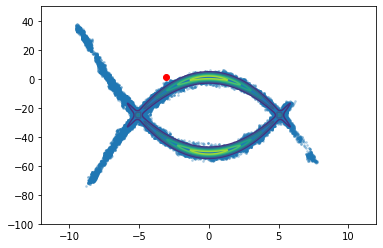}
\caption{DM Sampler}
\label{fig:dbdmfin}
\end{subfigure}
\begin{subfigure}{0.49\textwidth}
\includegraphics[width=0.9\linewidth, height=4cm]{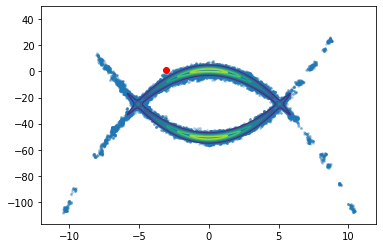} 
\caption{Scout MCMC}
\label{fig:dbscoutfin}
\end{subfigure}

\caption{Double Banana Samples from the finite variants of the DM Sampler and Scout MCMC. Both algorithms variants perform in line with their perpetually adapting counterparts.}
\label{fig:dbfinite}
\end{figure}

Finally, we plot the samples generated by the finite versions of the DM sampler and Scout MCMC in Figure \ref{fig:dbfinite}. The samples presented largely match those of the fully adaptive versions. This indicates that the bank of covariance matrices generated in the adaptive phase is sufficient to produce region specific samples as intended.

\subsection{Basis Vector Distribution}

As described briefly in the prelude to this section, the basis vector target consists of a series of normal distributions along the basis vectors in $\mathbb{R}^4$. Formally, the distribution is a mixture of the following normal distributions: 
$$
N\left(10 \mathbf{e}_1, \mathbf{I}_4 \right), N\left(-10 \mathbf{e}_1, \mathbf{I}_4 \right), N\left(10 \mathbf{e}_2, \mathbf{I}_4 \right), N\left(-10 \mathbf{e}_2, \mathbf{I}_4 \right),
$$
$$
N\left(10 \mathbf{e}_3, \mathbf{I}_4 \right), N\left(-10 \mathbf{e}_3, \mathbf{I}_4 \right), N\left(10 \mathbf{e}_4, \mathbf{I}_4 \right), \text{and } N\left(-10 \mathbf{e}_4, \mathbf{I}_4 \right)
$$
where $\mathbf{e}_i$ is a basis vector in the $i$th direction and $\mathbf{I}_4$ is the 4D identity matrix.

The key feature of this distribution is that the expected value is at the origin but there is negligible density there. As such, any MCMC algorithm that hopes to be successful must be able to cross a vast low probability desert to move between modes. Each algorithm was run for 40,000 iterations with the first 2,000 iterations discarded as burn-in. The results of the 4D basis vector target are presented in Table \ref{tab:basis}. 

Note that all of RWM, aRWM, MALA, and the DM sampler produced sample means that were far from the origin, the true mean. This indicates that these algorithms did not visit all of the modes in a balanced manner and likely got stuck in one or more select modes. This behaviour is not unexpected, however, as these algorithms have no mechanism to cross low probability boundaries. Scout MCMC and parallel tempering, in contrast, produced sample means approaching the origin with Scout MCMC outperforming both cases of parallel tempering. In this instance, there does not appear to be any major impact from increasing the number of chains from two to five in parallel tempering aside from a larger ESJD. Finally, we note that the finite variants of the DM sampler and Scout MCMC perform in line with their respective fully adapting versions, validating their specification as non-adapting approximations. 

\begin{figure}[ht]

\begin{subfigure}{0.49\textwidth}
\includegraphics[width=0.9\linewidth, height=4cm]{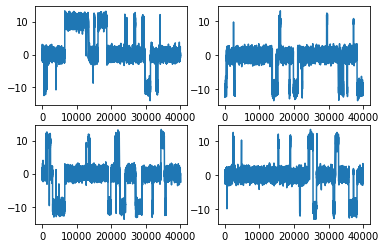}
\caption{PT (5 chains)}
\label{fig:ptbasis}
\end{subfigure}
\begin{subfigure}{0.49\textwidth}
\includegraphics[width=0.9\linewidth, height=4cm]{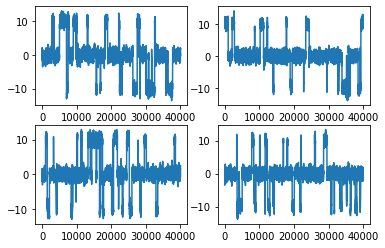} 
\caption{Scout MCMC}
\label{fig:scoutbasis}
\end{subfigure}

\caption{4D basis vector trace plots. Notice that both PT and Scout MCMC are able to reach -10, 0, and 10 in all four dimensions.}
\label{fig:basis}
\end{figure}

To confirm that all eight modes were visited (as opposed to, say, two opposing modes with a mean that is equal to the origin), we examine the trace plots of parallel tempering with 5 chains and Scout MCMC in Figure \ref{fig:basis}. Each trace plot represents one of the dimensions and for this target distribution, we should see the trace plots reaching -10, 0 and 10 in all dimensions. From the plots, it is clear that both algorithms are capable of moving between modes in a frequent manner and that all modes have been visited. It is at this juncture that we turn to the point of efficiency. Notice the difference in acceptance rate and ESJD between Scout MCMC and parallel tempering in Table \ref{tab:basis}. Scout MCMC has a tendency to accept almost twice as many proposals as parallel tempering even though parallel tempering takes larger steps. This is in part due to parallel tempering having no limit on the frequency of swap moves whereas Scout MCMC is set to only be able to consider swapping every 20 iterations.

\begin{center}
\begin{table}[ht]
\centering
 \begin{tabular}{ l | c c c c} 
 \hline
 & Accept (\%) &  $d$(E[X],$\hat{E}$[X]) & ESJD \\ [0.5ex] 
 \hline
 RWM & 37.88 & 10.03 & 1.09 \\ 
 \hline
 aRWM & 29.43 & 10.03 & 1.09 \\
 \hline
 PT (2 chains) & 37.63 & 2.67 & 1.47 \\
 \hline
 PT (5 chains) & 37.38 & 2.76 & 2.13 \\
 \hline
 MALA & 49.35 & 9.99 & 1.81 \\
 \hline
 DM Sampler & 70.89 & 10.09 & 0.39 \\
 \hline
 DM Finite & 69.72 & 10.05 & 0.39 \\
 \hline
 Scout MCMC & 70.60 & 1.01 & 1.01\\ 
 \hline
 Scout Finite & 71.06 &  1.26 & 1.04 \\ [1ex] 
 \hline
\end{tabular}
\caption{4D basis vector target results. Both versions of Scout MCMC and PT produced sample means that were closest to the true mean. PT slightly outperforms Scout MCMC in terms of ESJD but both variants of Scout MCMC have slightly closer means and higher acceptance rates.}
\label{tab:basis}
\end{table}
\end{center}

\subsection{Banana Bunch Distribution}

\begin{figure}[ht]
    \centering
    \includegraphics[height=4cm]{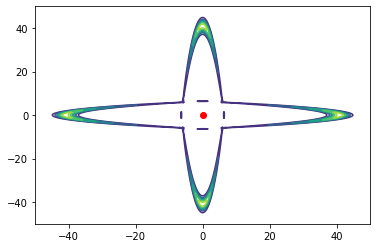}
    \caption{Projections of the banana bunch distribution on each pair of axes (x-y, x-z, and y-z). (Note: lighter contours indicate higher density, red dot indicates the origin)}
    \label{fig:plus}
\end{figure}

The next target consists of a mixture of 12 banana distributions in $\mathbb{R}^3$ arranged such that there is even less interaction than there is in the double banana example. We call this distribution the banana bunch. As this example is in $\mathbb{R}^3$, we cannot simply present the contours. However, the distribution can be understood as the mixture of three groups. The projection of the target on each pair of axes (x-y, x-z, and y-z) appears as the contours in Figure \ref{fig:plus}.

\begin{figure}[ht]
    \centering
    \includegraphics[height=4cm]{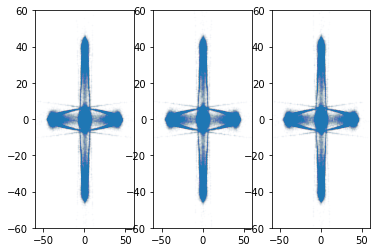}
    \caption{True banana bunch samples. Presented are samples generated directly from the target distribution projected onto the x-y, x-z, and y-z plane.}
    \label{fig:truth}
\end{figure}

Combining all three groups will result in the intersection of the apexes of two component distributions at $\pm 40$ along each axis. This results in a target with six modes, each far from the origin, and 24 tails extending from the modes towards each other. Once again, we capitalize on the symmetry of our targets and find the expected value to be at the origin. Though the origin has negligible density, scatter plots of samples projected down to planes composed of the basis vectors can be slightly misleading. Figure \ref{fig:truth} presents a sample generated directly from the target distribution. The points seemingly at the origin are actually "above" and "below" the origin with respect to the axis missing from the respective plot.

Given the more complex nature of this distribution, we increase the number of samples for all algorithms to 100,000 with the first 1,000 discarded as burn-in. In addition to acceptance rate, first moment, and ESJD, we also consider the second moment, the expectation of the element-wise square of the samples. Such an expectation will assess how well the tails have been explored. A sample that concentrates too heavily in the 6 modes will overshoot this expectation even if it successfully produces a mean near the origin. The true value of this expectation is $\begin{bmatrix} 400 & 400 & 400 \end{bmatrix}$. Table \ref{tab:3d} presents the results of our experiment.

\begin{center}
\begin{table}[ht]
\centering
 \begin{tabular}{ l | c c c c c} 
 \hline
 & Accept (\%) & $d$($E$[X],$\hat{E}$[X]) & $d$($E$[X$^2$],$\hat{E}$[X$^2$]) & ESJD\\ [0.5ex] 
 \hline
 RWM & 48.19 & 17.04 & 342.9 & 1.19\\ 
 \hline
 aRWM & 1.76 & 1.97 & 338.0 & 14.00\\
 \hline
 PT (2 chains) & 41.38 & 15.01 & 377.7 & 1.72\\
 \hline
 PT (5 chains) & 42.88 & 11.76 & 132.0 & 3.54\\
 \hline
 MALA & 3.17 & 12.62 & 576.7 & 0.1 \\
 \hline
 DM Sampler & 77.05 &  3.83 & 316.7 & 0.98\\
 \hline
 DM Finite & 57.71 &  12.05 & 334.0 & 0.77\\
 \hline
 Scout MCMC & 79.11 &  1.26 & 88.5 & 13.53\\ [1ex] 
 \hline
 Scout Finite & 63.34 &  2.6 & 109.6 & 10.99\\
 \hline
\end{tabular}
\caption{Banana bunch results. Although most algorithms come quite close to the true expected mean, not all are successful in finding the squared mean. The best performing ones are Scout MCMC, the finite variance of Scout MCMC, and PT with five chains. However, even with five chains, PT has a much less favourable ESJD than either Scout MCMC, each only utilizing two chains.}
\label{tab:3d}
\vspace{-20pt}
\end{table}
\end{center}

Perhaps the most surprising result in Table \ref{tab:3d} is that the sample mean produced by each algorithm is quite close to the true expected value although aRWM and Scout MCMC are clearly the best performers on that front. Regarding the squared expected value, Scout MCMC performs the best even though the results for parallel tempering with 5 chains presents a convincing case for its convergence properties. We note however that parallel tempering with 2 chains fails to match the performance of either Scout MCMC or parallel tempering with 5 chains. This suggests that 2 chains is not generally sufficient without a more nuanced strategy on the main chain such as using the DM sampler in Scout MCMC. Finally, with respect to ESJD, we note that aRWM and Scout MCMC are clearly the best performers using this efficiency metric. However, we warn the reader to recognize that aRWM accepted less than 2\% of proposals, did not produce a sample second moment that was close to the true second moment, and finished the algorithm with a covariance matrix with diagonal $\begin{bmatrix} 862 & 1000 & 209 \end{bmatrix}$. This is an indication that as the dimension increases, aRWM relies heavily on expanding its reach to cover the whole region of interest rather than conforming to the shape of the target distribution. The cost of this behaviour is that the proposals are not always of high quality and one must hope that it produces enough proposals to extract a decent set of good samples in a limited amount of time. In this case, the enormous proposal distribution was not sufficient to fully explore the tails leading in from the modes towards the origin, thus producing an inaccurate sample second moment. 

\begin{figure}[ht]

\begin{subfigure}{0.49\textwidth}
\includegraphics[width=0.9\linewidth, height=4cm]{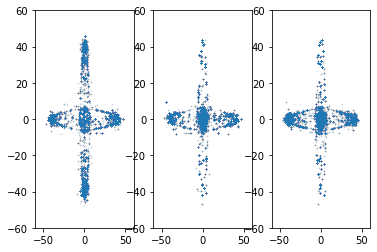} 
\caption{Adaptive RWM}
\label{fig:aproj}
\end{subfigure}
\begin{subfigure}{0.49\textwidth}
\includegraphics[width=0.9\linewidth, height=4cm]{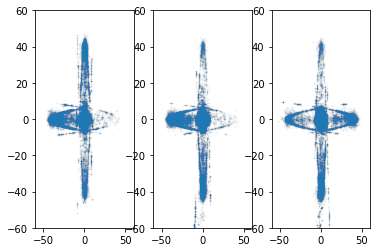} 
\caption{Parallel Tempering (5 chains)}
\label{fig:ptproj}
\end{subfigure}

\begin{subfigure}{0.49\textwidth}
\includegraphics[width=0.9\linewidth, height=4cm]{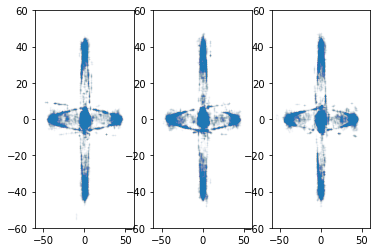}
\caption{Scout MCMC}
\label{fig:sproj}
\end{subfigure} 
\begin{subfigure}{0.49\textwidth}
\includegraphics[width=0.9\linewidth, height=4cm]{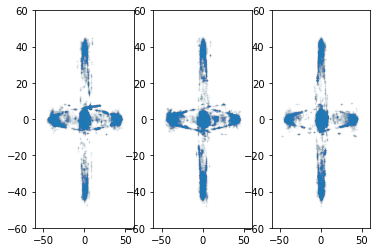}
\caption{Scout Finite}
\label{fig:sfproj}
\end{subfigure} 

\caption{Banana Bunch Samples (projected onto the x-y, x-z, and y-z planes). aRWM produces much sparser plots that reflect its inability to explore the tails of the bananas. PT, despite having more samples, still results in inaccurate sample statistics because of its imbalance, with many points skewing to one side. The two Scout MCMC variants produce the most optimal results with a large and balanced set of samples as well as strong exploration of the banana tails.}
\label{fig:projs}
\end{figure}

In order to be complete, however, we must also examine the plots of samples (projected to 2D) to understand whether our algorithms truly explore all modes and tails. Once again, we refer the reader to Figure \ref{fig:truth} to view the expected behaviour. Given that RWM, MALA, and the DM sampler have no mechanisms for inter-mode movement and that the numerical results reflect this fact, we focus on the remaining algorithms from here. We also focus on the 5 chain version of parallel tempering given that it outperformed the 2 chain version.

The visual results of Figure \ref{fig:projs} reflect the behaviours noted by the diagnostics in Table \ref{tab:3d}. Adaptive RWM struggles to explore the tails as well as the z-axis which results in the poor squared expected value. Parallel tempering, in contrast, performs better in the tails but does not explore the distribution evenly within the specified number of iterations which manifests in the poorer expected value. In addition, there are a number of samples well beyond the modes that are the result of a swap from a tempered chain to the main chain.  With Scout MCMC and its finite variant, we see the most appropriate distinction between mode and tail concentrations which manifests in the best squared expectation. They also do so with a higher frequency of points than aRWM or parallel tempering. The highly efficient proposals are realized even when the large distance between modes of the basis vector example is combined with the unusual geometry of the banana examples thus illustrating the ability of Scout MCMC to deliver on the promises of a multimodal sampler that excels at rapid adaptation to local geometry.

\subsection{Bayesian Linear Regression with Horseshoe Priors}

The final example target distribution is motivated by a more practical scenario. MCMC methods are frequently used in Bayesian inference and there is a lot of work being done to develop methods that can sample from complex posterior distributions. To test the DM sampler and Scout MCMC in such a situation, we construct a Bayesian linear regression scenario with half of the coefficients set to zero. In Bayesian settings, sparsity in regression coefficients can be induced using a horseshoe prior \citep{pmlr-v5-carvalho09a}. The horseshoe prior is known to be difficult to sample from given the use of the Cauchy distribution. Our example uses 20 independent variables and 1000 observations generated from a standard normal distribution. The true values of the first 10 coefficients are $\begin{bmatrix} 10 & 5 & 2.5 & 2 & 1 & 0.75 & 0.5 & 0.25 & 0.2 & 0.1\end{bmatrix}$ and the second set of 10 coefficients are all set to 0. In addition to the 20 coefficients, there is one error variance parameter, 20 horseshoe coefficient parameters, and one global horseshoe parameter for a total of 42 parameters to sample.

We execute each of the algorithms for 50,000 iterations with the first 5,000 discarded as burn-in. We modify some of the algorithms to accommodate the higher dimension of this example. For parallel tempering, we execute a 2 chain version as before but replace the 5 chain version with 10 chains. For the DM sampler, we adapt the step size to shrink upon rejecting a sample. Finally, for both the DM sampler and Scout MCMC, we only execute the adaptive versions. To compare across algorithms, we consider the acceptance rate, the distance between the true and estimated regression coefficients, and the ESJD. Table \ref{tab:horseshoe} presents the results of the experiment.

\begin{center}
\begin{table}[ht]
\centering
 \begin{tabular}{ l | c c c c c} 
 \hline
 & Accept (\%) & $d$($E$[X],$\hat{E}$[X]) & ESJD\\ [0.5ex] 
 \hline
 RWM & 0.48 & 0.21 & 1.0e-4\\ 
 \hline
 aRWM & 1.92 & 7.64 & 1.0e-2\\
 \hline
 PT (2 chains) & 0.52 & 0.2 & 5.7e-5\\
 \hline
 PT (10 chains) & 0.45 & 0.18 & 3.8e-5\\
 \hline
 MALA & 90.98 & 2.12 & 1.9e-4\\
 \hline
 DM Sampler & 34.98 & 0.23 & 2.4e-3\\
 \hline
 Scout MCMC & 5.51 & 0.26 & 2.1e-2\\ [1ex] 
 \hline
\end{tabular}
\caption{Bayesian linear regression with horseshoe priors results.}
\label{tab:horseshoe}
\vspace{-20pt}
\end{table}
\end{center}

As illustrated in the results, the majority of the algorithms tested produced posterior means that are close to the true coefficient values with aRWM being a notable exception. Also of interest are the ESJD values. Of the algorithms that did converge to the true parameter values, only the DM sampler and Scout MCMC had reasonably high ESJD values, suggesting that they sampled most effectively from the posterior. This example illustrates that both algorithms introduced here can be competitive with established MCMC methods in a practical setting.

We conclude this section by repeating the notion of efficient and effective sampling. We find that aRWM may be efficient from an ESJD perspective and parallel tempering is effective as a way to explore different modes, but neither prove to be adequate across the board. Instead, Scout MCMC proves itself as an efficient and effective "smart" sampler by adopting a strategy of combining rapid regional adaptation with heavy tempering for mode swapping. 

\section{Discussion and Future Work}

In this paper we have introduced an algorithm designed to rapidly adapt to the local behaviour of a given target distribution. Such adaptation is accomplished through the minimization of the information projection (I-projection) side of the KL Divergence between the target distribution and the proposal distribution family. By combining this Divergence Minimization sampler with one highly tempered chain to create Scout MCMC, we illustrate how the DM sampler may integrate into other existing MCMC approaches to combine algorithm strengths. Finally, we leverage the adaptation of the DM sampler and Scout MCMC in a two-stage algorithm that uses the produced covariance matrices of the DM sampler and Scout MCMC in the first phase to initialize a follow up Metropolis-Hastings phase that adheres to standard convergence criteria. This finite adaptation algorithm continues to use optimized local samplers to efficiently sample from local geometries without needing perpetual adaptive steps.

We have presented the baseline algorithm in this paper and believe that there is much room for future research. For example, the criteria required for a non-diminishing perpetually adaptive algorithm to converge remains under-explored. Moreover, one might be interested in studying whether there is an optimal frequency to adaptation and swapping or whether there are certain target geometries that are more or less challenging to explore. Smaller changes such as adapting step size and other fixed parameter inputs are also possibilities. Finally, there is certainly room to explore different objective functions in the DM sampler. Some possibilities include testing alternate similarity measures or replacing the regulating term to encourage different behaviours. Such modifications could further improve the performance of the DM sampler and Scout MCMC beyond what has been demonstrated in this paper.



\newpage

\appendix

\section{Appendix}

\subsection{Approximating the Gradient} \label{Gradient}

Since,
$$
\mathcal{J}(x) = \beta H_q + \beta E_\epsilon\left[\log p(x + L\epsilon)\right] + E_\epsilon\left[\min \left\{0, \log p(x + L\epsilon) - \log p(x) \right\}\right]
$$

the gradient of $\mathcal{J}(x)$ is:

$$
\begin{aligned}
&\nabla_L\mathcal{J}(x) \\
&= \nabla_L\beta H_q + \nabla_L\beta E_\epsilon\left[\log p(x + L\epsilon)\right] + \nabla_LE_\epsilon\left[\min \left\{0, \log p(x + L\epsilon) - \log p(x) \right\}\right] \\
&= \beta \nabla_L H_q + \beta  E_\epsilon\left[\nabla_L\log p(x + L\epsilon)\right] + E_\epsilon\left[\nabla_L\min \left\{0, \log p(x + L\epsilon) - \log p(x) \right\}\right]
\end{aligned}
$$

Consider each of the three terms of the above of $\nabla_L\mathcal{J}(x)$ individually:
\begin{enumerate}[align = left]
    \item[Term 1:] $\beta \nabla_L H_q$. We can use the form of the entropy of a multivariate normal distribution to evaluate this gradient:
    $$
    \begin{aligned}
    \beta \nabla_L H_q &= \beta \nabla_L\left(\frac{k}{2} \log (2\pi e) + \frac{1}{2} \log(|L||L^T|)\right) \\
    &= \beta \nabla_L\left(\frac{k}{2} \log (2\pi e) + \frac{1}{2} \sum_{i=1}^k\log L_{ii}^2\right) \\
    &= \beta \nabla_L\left(\sum_{i=1}^k\log L_{ii}\right) \\
    &= \beta \text{diag}\left(\frac{1}{L_{11}},\dots,\frac{1}{L_{kk}}\right)
    \end{aligned}
    $$
    \item[Term 2:] $\beta E_\epsilon\left[\nabla_L\log p(x + L\epsilon)\right]$. First, note the following:
    $$
    \begin{aligned}
    \beta E_\epsilon\left[\nabla_L\log p(x + L\epsilon)\right] &= \beta E_\epsilon\left[\frac{1}{p(x + L\epsilon)}p'(x + L\epsilon)\epsilon^T \right]
    \end{aligned}
    $$
    The expectation on the right-hand side does not simplify cleanly. However, the interior of the expectation is simple enough to evaluate for a given $\epsilon$. As such we can draw a number of $\epsilon_j \sim N(0,1)$ at each iteration and compute an unbiased estimate of $\beta E_\epsilon\left[\nabla_L\log p(x + L\epsilon)\right]$ with Simple Monte Carlo. That is, at each iteration, we compute:
    $$
    \beta E_\epsilon\left[\nabla_L\log p(x + L\epsilon)\right] \approx \frac{1}{J} \sum_{j=1}^J\frac{\beta}{p(x + L\epsilon_j)}p'(x + L\epsilon_j)\epsilon_j^T 
    $$
    \item[Term 3:] $E_\epsilon\left[\nabla_L\min \left\{0, \log p(x + L\epsilon) - \log p(x) \right\}\right]$. Similar to the second piece of the gradient, note that this expectation does not simplify but we can produce an unbiased estimate by relying on a series of draws of $\epsilon_j \sim N(0,1)$ in a given iteration. 
    $$
    \begin{aligned}
    &E_\epsilon\left[\nabla_L\min \left\{0, \log p(x + L\epsilon) - \log p(x) \right\}\right] \\ &\approx \frac{1}{J}\sum_{j=1}^J\nabla_L\min \left\{0, \log p(x + L\epsilon_j) - \log p(x) \right\}
    \end{aligned}
    $$
    However, the presence of the minimum operator suggests this summation will not simplify in the same way as the previous component. Considering the two cases, we can naturally separate them depending on if $\log p(x + L\epsilon_t) \ge \log p(x)$.
    
    In the first case, if $\log p(x + L\epsilon_t) \ge \log p(x) $, acceptance of the proposal under a Metropolis framework is guaranteed and:
    $$
    \nabla_L\min \left\{0, \log p(x + L\epsilon_t)  - \log p(x)\right\} = 0
    $$
    If $\log p(x + L\epsilon_t) < \log p(x)$, then the Metropolis ratio is less than 1 and we have,
    $$
    \begin{aligned}
    &\nabla_L\min \left\{0, \log p(x + L\epsilon_t) - \log p(x) \right\} \\
    &= \nabla_L \left(\log p(x + L\epsilon_t) - \log p(x)\right) \\
    &= \frac{1}{p(x+L\epsilon_t)}p'(x+L\epsilon_t)\epsilon_t^T
    \end{aligned}
    $$
\end{enumerate}

Consolidating the three terms, to search for an optimal local proposal distribution, at each iteration of the MCMC chain we perform the following gradient-based update (we omit the iteration subscript $t$ on $x$ and $L$ for clarity purposes):

$L_{t+1} = L_t + \gamma\nabla_L\mathcal{J}(x)$
where 
$$
\begin{aligned}
\nabla_L\mathcal{J}(x) &= \beta \text{diag}\left(\frac{1}{L_{11}},\dots,\frac{1}{L_{kk}}\right) + \frac{1}{J}\sum_{j=1}^J\frac{\beta}{p(x + L\epsilon_j)}p'(x + L\epsilon_j)\epsilon_j^T + \\
&\quad \frac{1}{J}\sum_{j=1}^J\nabla_L\min \left\{0, \log p(x + L\epsilon_j) - \log p(x) \right\}.
\end{aligned} 
$$
Note further that the interior of the Term 3 summation reduces into the following two cases depending on the value of $\epsilon_j$,

$$
\begin{aligned}
&\quad\nabla_L\min \left\{0, \log p(x + L\epsilon_j) - \log p(x) \right\} \\
&=\begin{cases} 0&\mbox{if } \log p(x + L\epsilon_j) \ge \log p(x) \\
\frac{1}{p(x + L\epsilon_j)}p'(x + L\epsilon_j)\epsilon_j^T & \mbox{if } \log p(x + L\epsilon_j) < \log p(x) \end{cases}.
\end{aligned}
$$

Note that the current position is denoted $x$, the proposal is $y= x + L\epsilon$, the standard multivariate draw is $\epsilon_j$, and $\gamma$ is the predetermined step size.

\clearpage
\bibliographystyle{abbrvnat}
\bibliography{refs}

\end{document}